\documentclass{article}  
\usepackage{graphicx, subfig, caption}
\usepackage{natbib}
\title{The Orbital Period of Three Cataclysmic Variables from WASP Data}
\author{Patrick Wils$^1$\\
\small
 $^1$Vereniging Voor Sterrenkunde, Belgium;  
email: patrickwils@yahoo.com \\
      }
\date{October 2010}

\begin{document}
\maketitle 

\begin{abstract}
The publicly available WASP data are analysed to determine the orbital periods of the cataclysmic variables V378~Peg, SDSS~J171456.78+585128.3 and ASAS~150946-2147.7.
\end{abstract}

\section{Introduction}

The exoplanet transit survey WASP (Wide Angle Search for Planets) has been taking wide field images from 2004
using two instruments located in La Palma and South Africa.
In the first public data release of the WASP archive all the light curve data and images from 2004 up to 2008 
from both the Northern and Southern hemispheres instruments were made available \citep{WASP}. 
Since these light curves are used to search for exoplanet transits, they can also be used to study other low amplitude variability in stars.
\citet{Thomas} successfully used the data to study the orbital period variations of three cataclysmic variables (CVs).
In this study the publicly available WASP data will be used to determine the orbital periods of three other CVs.
In the analysis only TAMUZ \citep{Tamuz} corrected data were used for which the uncertainty on the magnitude is less than 0.1.

\section{V378~Peg}

V378~Peg = PG~2337+300 has not been studied since its classification by \citet{Koen} as a cataclysmic variable.
The X-ray source 1RXS~J234002.7+301808 is probably also related.
\citet{Koen} observed irregular brightness variations of up to 0.3 magnitude with a timescale of a few minutes over a three hour period.

\begin{figure*}
\centering
\includegraphics[width=12cm]{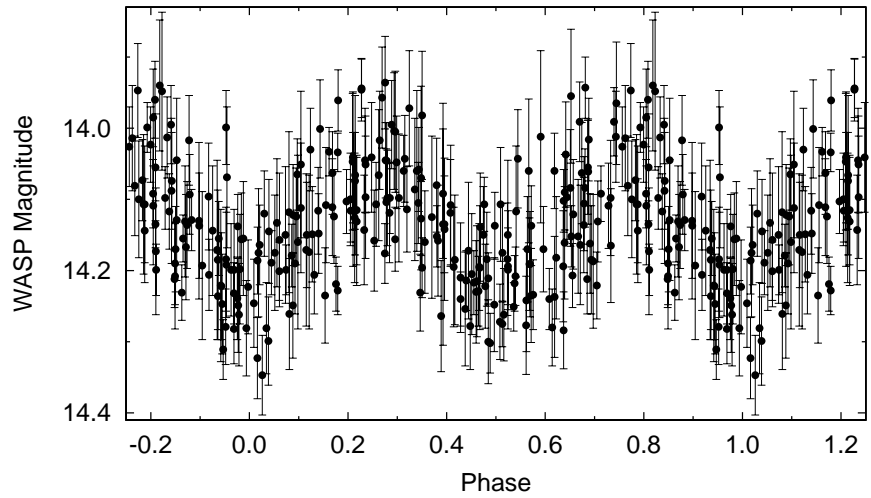}
\caption{Phase plot of 10-point averages of WASP data of V378~Peg, using a period of 0.26985 days.
The error bars represent the standard deviation of the 10-point averages.}
\label{V378Peg}
\end{figure*}

The WASP data (1SWASP~J234004.30+301747.5) show low amplitude variations around magnitude 14.1.
A period search using the Phase Dispersion Minimization (PDM) technique of \citet{Stellingwerf} 
shows a sinusoidal light curve with a period of 0.1349 days, with an average amplitude of around 0.15 magnitude, 
but with a lot of scatter likely caused in part by the irregular variations seen by \citet{Koen}.
The 1-day alias of 0.1560 days cannot be entirely excluded to be the real period, 
but the shorter value seems to be a better fit for the long wave in the light curve of \citet{Koen}.
This variation may be caused by the rotation of the hot spot or the irradiation of the cool companion, 
but it is also possible that the orbital period is twice the given value, 
in which case the variation is caused by the ellipsoidal shape of the red dwarf star.
Fig.~\ref{V378Peg} shows the phase plot using a period of 0.26985 days, but the phase plot of the longer period is almost indistinghuisable.
Neither can a distinction be made between the minima or maxima in the double period solution.
The available photometry from 2MASS \citep{2MASS} and the Galaxy Evolution Explorer \citep[GALEX;][]{GALEX} 
are compatible with a single object with a black body temperature of $12000K$ or higher and do not show the presence of a cool companion.
Hence it must be too cool to be detected, and therefore cannot be responsible for the variations seen in the light curve.
In that case an orbital period of 0.1349 days is the most likely solution.

\section{SDSS~J171456.78+585128.3}

GUVV-2~J171456.8+585128.3 was discovered to be variable by \citet{Wheatley} in data from the GALEX satellite, 
and as SDSS~J171456.78+585128.3 = 1RXS~J171456.2+585130 it was found to be a CV by \citet{Agueros}.
The latter authors found the spectrum to be that of a $K4$ star with an invisible companion.  
Their sparse radial velocity measurements suggested an orbital period of $\sim10$ hours.
The WASP data for this object (1SWASP~J171456.79+585128.6) show variations around magnitude 14.6 
with an amplitude of 0.2 magnitude and a period of about twice the value proposed by \citet{Agueros}. 
The data follow the ephemeris:

\begin{equation} 
\label{EphGUVV} 
   HJD~Min~I = 2453260.118 + 0.83803 \times E 
\end{equation}

\begin{figure*}
\centering
\includegraphics[width=12cm]{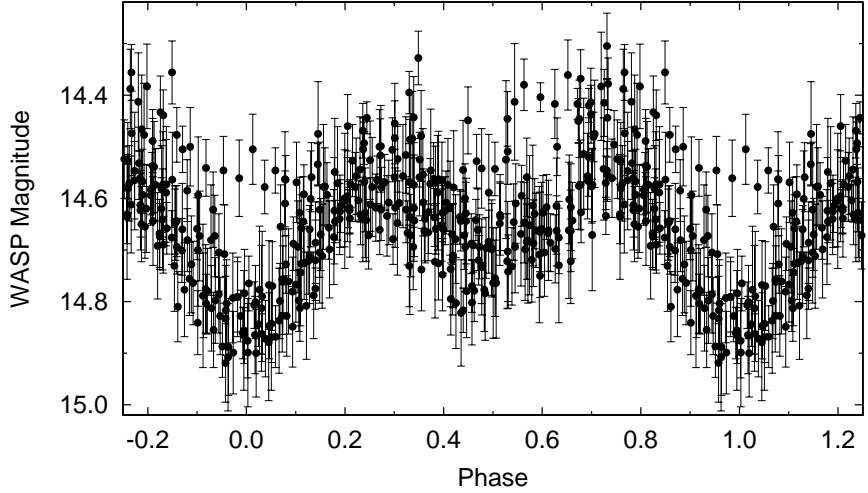}
\caption{Phase plot of 10-point averages of WASP data of GUVV-2~J171456.8+585128.3, using the ephemeris given in Eq.\ref{EphGUVV}.  
The error bars represent the standard deviation of the 10-point averages.}
\label{GUVV-2J171456.8+585128.3}
\end{figure*}

The light curve in Fig.~\ref{GUVV-2J171456.8+585128.3} shows a secondary minimum.
These variations are very likely caused by the ellipsoidal shape of the $K4$ star and irradiation or limb darkening effects to account for the different brightness of the minima. 
The data from the Northern Sky Variability Survey \citep[NSVS;][]{Wozniak} 
are compatible with the orbital period found here \citep{cvnet}.

\section{ASAS~150946-2147.7}

This object was suspected to be a dwarf nova in outburst by \citet{Pojmanski} in data from the All-Sky Automated Survey \citep[ASAS;][]{ASAS}.  
They noted that the object is a blend between two objects.
\citet{Henden} identified the outbursting object to be identical to 2MASS~J15094657-2147462, the brighter of the two objects.
Therefore the influence of the fainter object on the brightness of the variable is negligable.
According to \citet{Uemura} ASAS~150946-2147.7 is a candidate black hole X-ray binary based on data from the Swift satellite.

\begin{figure*}
\centering
\includegraphics[width=12cm]{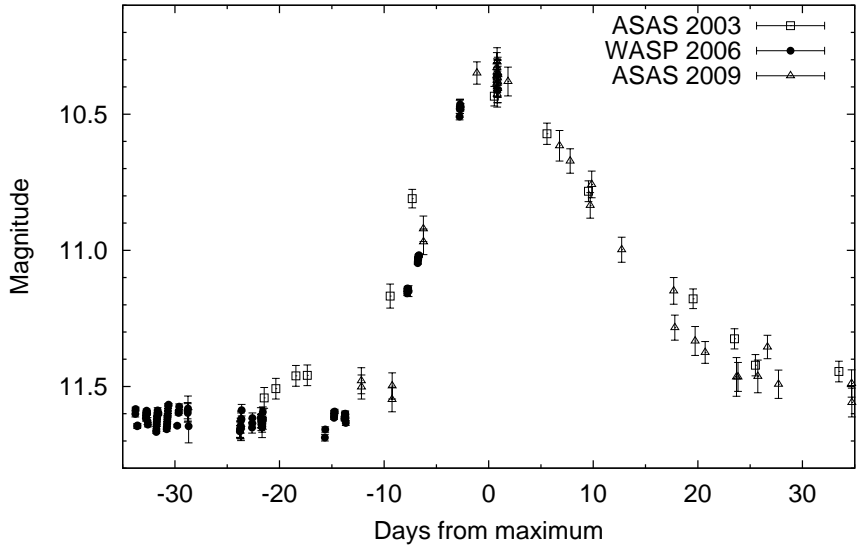}
\caption{Comparison of the profiles of the three observed outbursts of ASAS~150946-2147.7.  
Two outbursts were observed in the $V$ band by ASAS \citep{ASAS}, and one was observed unfiltered by WASP \citep{WASP}.  
The WASP data are plotted as averages of 10 consecutive points. }
\label{outburst}
\end{figure*}

Two outbursts were observed by ASAS, in August 2003 and April 2009.  
The WASP archive (object 1SWASP~J150946.56-214746.5) contains the rising branch of another outburst in July 2006, 
matching the shape of the ASAS outbursts (see Fig.~\ref{outburst}).
The total outburst amplitude reached just over one magnitude in all cases.
These outbursts last less than a month with a slow rise to maximum, taking almost as long as the fade to minimum.  
This behaviour is rather atypical for a dwarf nova and is normally only seen in CVs with an orbital period longer than one day, 
such as the old nova GK~Per \citep[orbital period 1.9968d;][]{Crampton}, V630~Cas \citep[2.5639d;][]{Orosz} 
and SDSS~J204448.92-045928.8 \citep[1.68d;][]{Peters}.
Recent outbursts of these dwarf novae have been described by \citet{Evans}, \citet{Shears} and \citet{Wils} respectively.

\begin{figure*}
\centering
\includegraphics[width=12cm]{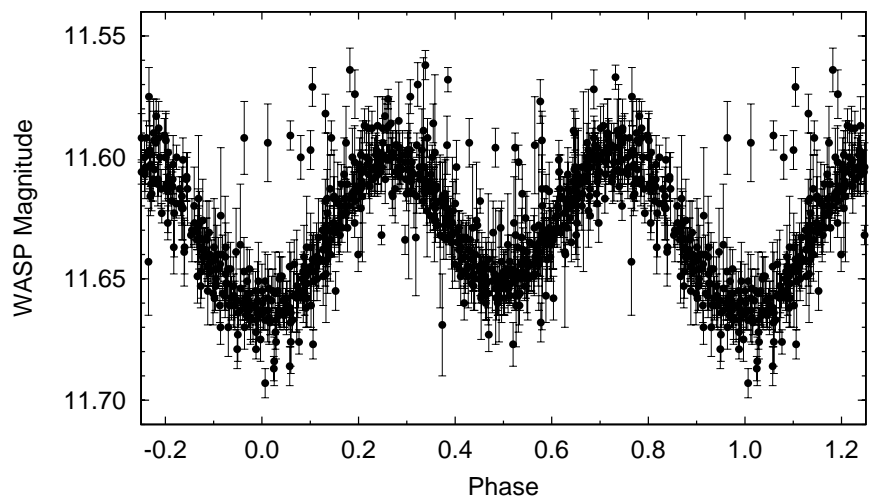}
\caption{Phase plot of 10-point averages of WASP data of ASAS~150946-2147.7, using the ephemeris given in Eq.\ref{EphASAS}.  
Only data in quiescence were used.
The error bars represent the standard deviation of the 10-point averages.}
\label{ASAS150946-2147.7}
\end{figure*}

\citet{Pojmanski} further detected weak 0.1 mag modulations in the ASAS-3 data at quiescence, with a period of 0.351213 or 0.206187 days (these are 1-day aliases of each other).
Because of the higher cadence of the WASP observations they are better suited to distinguish between aliases in this case.
A period search of the out-of-outburst data unambiguously revealed the period to be 0.70242 days, exactly twice the longer period found by \citet{Pojmanski}.  
The light curve in Fig.~\ref{ASAS150946-2147.7} shows a double wave varying from magnitude 11.60 to 11.67 
with one minimum fainter than the other (secondary minimum at magnitude 11.65), 
This is likely caused by the changing aspects of an ellipsoidal star during an orbital revolution.
An ephemeris for the primary minimum has been calculated as follows:

\begin{equation} 
\label{EphASAS} 
   HJD~Min~I = 2453880.235 + 0.70242 \times E 
\end{equation}

No phase shifts or period changes were detected after the outburst.

\section{Conclusion}

The orbital period of V378~Peg cannot be unambiguously determined from WASP data.  
It is either 0.1349 or 0.1560 days, or twice these values, but the first value is the preferred solution.
For SDSS~J171456.78+585128.3 the orbital period was found to be 0.83803 days with two minima and two maxima per orbit caused by the ellipsoidal shape of the red dwarf.
The orbital period of ASAS~150946-2147.7 was found to be 0.70242 days, again with two minima and maxima per orbit.  
Although not as long as the orbital period of GK~Per, the outburst behaviour of ASAS~150946-2147.7 is similar. 
Further spectroscopic study should reveal its true nature.
\\

\noindent{\bf Acknowledgements}\\
We have used data from the WASP public archive in this research. 
The WASP consortium comprises of the University of Cambridge, Keele University, University of Leicester, The Open University, 
The Queen's University Belfast, St. Andrews University and the Isaac Newton Group. 
Funding for WASP comes from the consortium universities and from the UK's Science and Technology Facilities Council.

This study made use of NASA's Astrophysics Data System, and the SIMBAD and VizieR 
databases operated at the Centre de Donn\'ees astronomiques (Strasbourg) in France.

\end{document}